\documentstyle[aps,preprint,epsf]{revtex}
\tightenlines

\begin{document}

\title{Microscopic Nuclear Level Densities from
 Fe to Ge  by the Shell Model Monte Carlo Method}

\author{H. Nakada$^1$ and Y. Alhassid$^2$}

\address{$^1$Department of Physics, Chiba University,
Inage, Chiba 263-8522, Japan \\
$^2$Center for Theoretical Physics, Sloane Physics Laboratory,
Yale University, New Haven, Connecticut  06520, U.S.A.}

\date{\today}

\maketitle

\begin{abstract}

We calculate microscopically total and parity-projected level
densities for $\beta$-stable even-even nuclei between  Fe and Ge,
using the shell model Monte Carlo methods in the complete
$(pf+0g_{9/2})$-shell. A single-particle level density parameter $a$
and backshift parameter $\Delta$ are extracted by fitting the
calculated densities to a backshifted Bethe formula,
and their systematics are studied across the region.
Shell effects are observed in $\Delta$ for nuclei with $Z=28$ or
$N=28$ and in the behavior of $A/a$ as a function of the number of
neutrons. We find a significant parity-dependence of the level
densities for nuclei with $A \alt 60$, which diminishes
as $A$ increases.

\end{abstract}


Nuclear level densities are important in theoretical estimates
of various compound nuclear reactions.
For example,  neutron-capture reaction rates are approximately
proportional to the corresponding level densities in the neutron
resonance region. Neutron capture rates are relevant to the $s$- and
 $r$-processes in nucleosynthesis.
The $s$-process  proceeds by a sequence of neutron-capture
 reactions along the $\beta$-stable line,
since the  $\beta$-decay of a radioactive nucleus will occur much
 faster  than the capture of a neutron.  
The waiting points of the $r$-process
are determined by the balance between the rates of neutron-capture
 and of photoejection of a neutron.
 Accurate methods for calculating  level densities are
  especially useful for nuclei where neutron-capture cross-sections
 are unavailable experimentally.
Nuclei in the iron-group are of special interest since they are
 the seed nuclei for the synthesis of the heavy elements by the $s$-
and $r$-processes.

Experimentally, level densities are usually extracted
from direct counting at low excitation energies ($E_x \alt 5$ MeV),
 from neutron or proton resonance data~\cite{ref:Dilg}, from
charged particles spectra at intermediate energies
($5 \alt E_x  \alt 15$ MeV)~\cite{ref:Hui,ref:LVH},
 and from Ericson fluctuation analysis at higher  energies
 ($E_x \agt 15$ MeV)~\cite{ref:Hui}.
  In many nuclei, the densities at low energies
 and around the neutron resonance
 region are well-fitted to phenomenological modifications
 of the Fermi gas model~\cite{ref:BM1},
 in particular to the backshifted Bethe
formula (BBF)~\cite{ref:HWFZ,ref:CTT}.
However, the fitted single-particle level density parameter
$a$ and backshift parameter $\Delta$ are nucleus-dependent  and
 difficult to derive theoretically.  Consequently, level densities
 used in nuclear astrophysical studies are based on
 various empirical formulae for
$a$ and $\Delta$~\cite{ref:HWFZ,ref:CTT,ref:WFHZ,ref:RTK97}.
  Recently we  introduced a novel method to calculate microscopically
total and parity-projected  level densities~\cite{ref:NA97}
 in the framework of the auxiliary-field Monte Carlo method.
The method takes into account exactly correlations due to
 effective two-body interactions in a finite shell model space.
We showed that  realistic nuclear level densities can be calculated
  with good accuracy  if a sufficiently large model space is used.
 In particular,  we  applied the method to $^{56}$Fe and found
 the calculated total level  density to be in good agreement with the
 experimental one.  We found a significant parity-dependence of the
 level density, contrary to the common assumption in
astrophysics calculations that positive- and
 negative-parity levels have equal densities
 around the neutron resonance region.
Shell (or subshell) structure affects
 the parity dependence  of level densities,
as pointed out already in Ref.~\cite{ref:G88}
 for non-interacting fermions.
In this letter, we investigate the systematics
of the total and parity-projected level densities of
$\beta$-stable even-even nuclei between Fe and Ge.  The calculated
 level densities are found to be well-described by the BBF.
We extract both $a$ and $\Delta$ from the  microscopic Monte Carlo
 densities by fitting them to the BBF, and find shell effects
 in their systematics.
The  extracted parameters are also compared with empirical
values  and  with experimentally extracted values when available.

 Approximate methods to calculate level densities
 beyond the Fermi gas model were developed in the framework of
 the spectral averaging theory~\cite{ref:MF75,ref:SG88,ref:KM96}.
Level densities for several $pf$-shell nuclei were recently
 calculated~\cite{ref:KM96} 
by decomposing the level density into a
sum of non-interacting level densities convoluted with Gaussians
(the Gaussians reflect the effect of interactions).
Since  Gaussian spreading is justified only above a certain
 reference energy, the method used in Ref.~\cite{ref:KM96} requires
the knowledge of a complete set of levels
 up to sufficiently high energy.
While reasonable agreement with data  is found,
two parameters, both of which depend on the nucleus under study,
 were fitted to the level density data:
a reference energy and an interaction parameter.
In our  level density calculations, 
no adjustable parameters have been introduced,
yet the calculated level densities are in good agreement
 with the data.

  Neutron resonance energies  for $50\alt A\alt 70$ nuclei are
 in the range $E_x\alt 15$~MeV. To include the important excitations
 in this mass and energy region in a shell model approach, 
it is necessary to use the full $(pf+0g_{9/2})$ model space.
This model space is too large for conventional diagonalization
methods: instead we use the shell model Monte Carlo (SMMC)
 method~\cite{ref:SMMC}
 in the complete $pf$- and $0g_{9/2}$-shell.
For level density calculations we need to solve the interacting
shell model problem at finite temperature. The SMMC, based on an
 auxiliary-fields representation of the canonical density matrix
 at finite temperature, is particularly suitable for such calculations.
In this letter we discuss only even-even nuclei. Odd-$A$ and
 odd-odd nuclei will be considered in a future publication.   For an
appropriately chosen interaction, even-even nuclei are not subject
 to the Monte Carlo  sign problem, and  thus their level densities can
be calculated more accurately.
Except for even-odd effects due to the pairing correlations
 (which are reflected in the backshift parameter $\Delta$),
important features of   the level density systematics can  already be
 deduced from the study of even-even nuclei.

We use the isoscalar Hamiltonian discussed in Ref.~\cite{ref:NA97}.
The single-particle energies  are calculated in
 a Woods-Saxon potential  plus spin-orbit interaction \cite{ref:BM1}.
The two-body residual interaction includes monopole isovector pairing
whose strength is determined from experimental
 odd-even mass differences,
and a self-consistent surface-peaked interaction~\cite{ref:ABDK}.
 The latter is expanded in
quadrupole, octupole and hexadecupole terms that are appropriately
 renormalized. The Hamiltonian is uniquely determined for each
 nucleus by the above procedure,
with no adjustable parameters remaining.
Our Hamiltonian includes the dominating collective
 components of realistic effective interactions~\cite{DZ96},  yet
  has  the advantage that it satisfies the modified
 sign rule~\cite{ref:ADK}, and therefore has a good Monte Carlo sign
for even-even nuclei. This enables us to perform accurate Monte Carlo
calculations down to temperatures that are low enough to extract
 reliable ground state energies.   The method used to extract the
 total level density is explained in Ref.~\cite{ref:NA97}.
The canonical  thermal  energy $\langle H\rangle_\beta$
 (at a fixed proton and neutron number) is calculated in SMMC
as a function of inverse temperature $\beta$, and then integrated
 to find the canonical partition function $Z(\beta)$. 
The level density $\rho$  is evaluated in the  saddle-point 
approximation
 \begin{eqnarray}\label{7}
  \rho(E) = (2\pi \beta^{-2} C)^{-1/2}  e^{S} \;,
\end{eqnarray}
  in terms of the canonical entropy
 $S(E) = \beta E +  \ln Z $ and the heat capacity
$C  = - \beta^2 dE / d\beta$.
Here the relation between energy and
 temperature is determined by the
saddle-point condition $E(\beta) = \langle H\rangle_\beta$.
The parity-dependence of the level density is calculated  using  the
parity-projection method described in Ref.~\cite{ref:NA97}.

\begin{figure}
\epsfysize=6.0 cm
\centerline{\epsffile{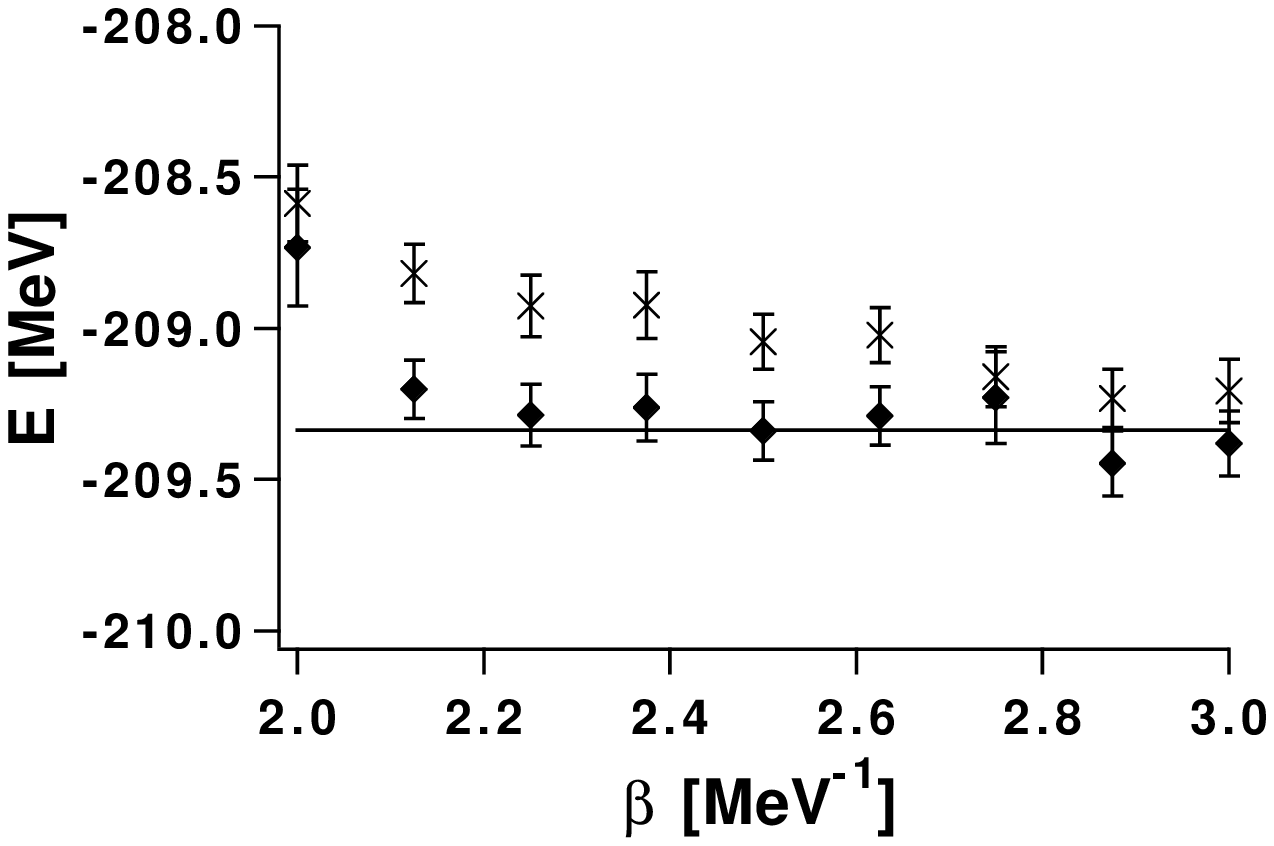}}
\vspace{2mm}
\caption{Thermal energy $E(\beta)$ in SMMC (crosses) and 
 ground state energy $E_0$ (diamonds) for   $^{58}$Fe
extracted via Eqs.~(\protect\ref{extrap_E}) and 
 (\protect\ref{extrap_J}).
The solid line is the average of the values for $E_0$
between  $\beta=2.5$ to $3~{\rm MeV}^{-1}$.}
\label{fig:extrap}
\end{figure}

\begin{figure}
\epsfysize=5.0 cm
\centerline{\epsffile{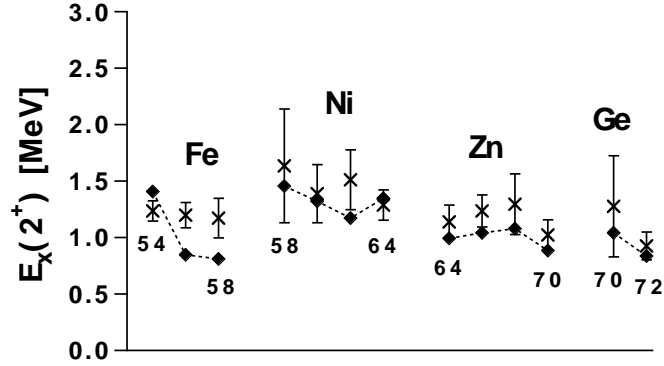}}
\vspace{2mm}
\caption{Excitation energies $E_x(2^+_1)$
 of the first $2^+$ state: comparison between the SMMC values
(calculated from Eq.~(\protect\ref{extrap_J})) (crosses)
and the experimental  values (diamonds).
Mass numbers are shown for several nuclei below the symbols.}
\label{fig:Ex(2+)}
\end{figure}

To compare with experimental data, it is necessary to find
 $\rho$ as a function of the excitation energy.
It is thus important to determine accurately
the ground-state energy $E_0$:  any error in $E_0$ will directly
 affect the value of the backshift parameter.  The ground state
 energy can be obtained from the thermal energy in the limit  of zero
 temperature, i.e.  $E(\beta\rightarrow\infty)$.  Since
 the auxiliary-fields propagator matrix  used in the Monte Carlo method
becomes  ill-conditioned \cite{ref:LG92} at  large
 $\beta$ ($\agt 5$~MeV$^{-1}$),
we have to determine the ground state energy from finite $\beta$
calculations,
where the lowest excited states still contribute to the thermal energy.
 The ground state of  even-even nuclei is $J^P=0^+$,
while the first excited state of almost all even-even nuclei is
 $J^P=2^+$ with an excitation energy of $\sim 1$ MeV
in the $50 \alt A \alt 70$ mass region.
For sufficiently low temperatures ($\beta \agt 2 - 3$ MeV$^{-1}$),
 the main contribution to  $E(\beta) - E_0$ arises
 from  thermal excitation to the $2^+_1$ state.
In the two-state model of $0^+_1$ and $2^+_1$, we have
\begin{eqnarray}\label{extrap_E}
  E(\beta) \approx E_0 + \eta E_x(2^+_1) ,
\end{eqnarray}
where   $E_x(2^+_1)$ is the excitation energy of the $2^+_1$
and  $\eta\equiv 1/\{1+\exp[\beta E_x(2^+_1)]/5\}$  (where the factor 5
 accounts for the spin degeneracy  of the $2^+$ state).
The ground state energy $E_0$ can then be extracted 
by a two-parameter fit of (\ref{extrap_E}) to the large-$\beta$ 
Monte Carlo data for $E(\beta)$.
 However, in SMMC we can also calculate
$\langle \vec{J}^2\rangle_\beta$, the
canonical expectation value of $\vec J^2$,
where $\vec{J}$ is the angular-momentum operator.
In the two-state model
\begin{eqnarray} \label{extrap_J}
   \langle \vec{J}^2 \rangle_\beta &\approx& 6 \eta \;.
\end{eqnarray}
 By fitting  the Monte Carlo data of
  $\langle \vec{J}^2\rangle_\beta$ to (\ref{extrap_J}),
 we can first extract $E_x(2^+_1)$, and then use this value
 in (\ref{extrap_E})  to determine the ground state energy $E_0$
 more accurately by a one-parameter  fit.
In Fig.~\ref{fig:extrap}, the extracted values of $E_0$
are shown for $^{58}$Fe as a function of $\beta$.
The present procedure
gives a stable $E_0$ value beyond $\beta=2.5~{\rm MeV}^{-1}$.
We adopt the $E_0$ and $E_x(2^+_1)$ values
by averaging  the values in the range $\beta=2.5$ to $3~{\rm MeV}^{-1}$.
Our method  not only provides a more
 accurate determination of  $E_0$ but also gives
$E_x(2^+_1)$, although with less accuracy
 (typically $\sim 0.1 - 0.2 $~MeV) than
 the ground state energy ($\sim 0.05$~MeV).
In Fig.~\ref{fig:Ex(2+)} we compare the calculated excitation energies
 with the experimental data, and a fairly good agreement is observed.
This is another confirmation that our  present interaction
includes properly  the dominating collective features of the
 realistic nuclear  force.
We remark  that the  first observed excited state
in $^{72}$Ge is a $0^+$,  an exception to the usual $2^+$.
It is not clear whether
our present Hamiltonian can reproduce this $0^+_2$ state.
Experimentally, the $2^+_1$ state  lies only $0.14$~MeV above
the $0^+_2$.  Because of the spin degeneracy factor,
 the thermal weight of  the $2^+_1$ is still about three times
larger than that of $0^+_2$  for $\beta=3~{\rm MeV}^{-1}$.
Hence we can neglect the contribution of 
 this $0^+_2$ in the above procedure.
The reliability of our Hamiltonian was  further  confirmed
by the average energy of the mass quadrupole excitation
in $^{56}$Fe~\cite{ref:NA97}.

\begin{figure}
\epsfysize=8.0 cm
\centerline{\epsffile{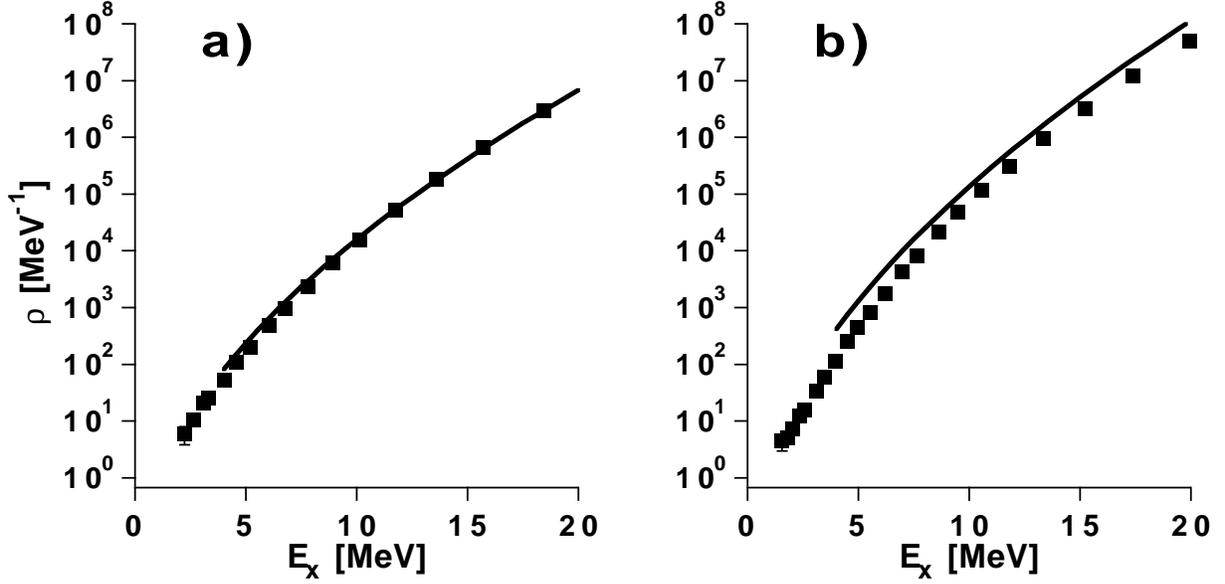}}
\vspace{2mm}
\caption{Total  level densities: comparison between SMMC
(solid squares with error bars)
 and experimentally determined level densities (solid lines).
a) $^{60}$Ni. The experimental BBF parameters
are taken from Ref.~\protect\cite{ref:LVH}.
b) $^{68}$Zn. The experimental BBF parameters
are taken from Ref.~\protect\cite{ref:Dilg}
 assuming rigid-body moment-of-inertia.}
\label{fig:rho_t}
\end{figure}

The BBF level density~\footnote{
The density given by Eq.  (\ref{BBF})  is
 sometimes referred to in the literature as ``state density''
 where each level with spin $J$ is weighted by a factor of 
 $2J +1$ (to include its magnetic
quantum number degeneracy).  All densities calculated in this paper
 are state densities.} at excitation energy $E_x$  is given by
\begin{eqnarray}\label{BBF}
 \rho (E_x) \approx
g  {{\sqrt\pi}\over{24}} a^{-\frac{1}{4}} (E_x - \Delta)^{-\frac{5}{4}}
 e^{2\sqrt{a (E_x - \Delta)}}
\end{eqnarray}
with $g=2$ for the total level density and $g=1$ 
for the parity-projected densities.
The SMMC total level densities are well described by (\ref{BBF}),
 and are in good agreement  with the level densities that are 
reconstructed from experimentally determined $a$ and $\Delta$.
For example, in  Fig.~\ref{fig:rho_t}
 we compare the SMMC total  level densities of
 $^{60}$Ni and $^{68}$Zn with the experimental ones
\cite{ref:Dilg,ref:LVH} and find good agreement, similar to that
 found  in  Ref.~\cite{ref:NA97}  for  the $^{56}$Fe level density.
The calculated parity-projected level densities
are also well fitted to the BBF  if parity-specific values for $a$ and
$\Delta$ (denoted by $a_\pm$ and $\Delta_\pm$) are  used.
In general these values differ from
those for the total level densities.

We  have extracted  the level density parameters
$a$ and $\Delta$ from SMMC level density  calculations
 for  a number of $\beta$-stable even-even nuclei
in the $50\alt A\alt 70$ region:
$^{54-58}$Fe, $^{58-64}$Ni, $^{64-70}$Zn and $^{70,72}$Ge.
In some of the previous analyses of  level densities
(see e.g. Ref.~\cite{ref:WFHZ}),
a simple empirical function was assumed for the backshift
 parameter $\Delta$, and the single-particle level density parameter $a$
was then fitted to the experimental data. In Ref.~\cite{ref:Dilg}
 both $a$ and $\Delta$ are fitted to the experimental data.
Empirical formulae for
 $a$ were proposed in Refs.~\cite{ref:Dilg,ref:WFHZ},
but different formulae yield quite different level densities.
Furthermore, there still remains a  discrepancy
between the experimental and the various empirical values of $a$.
While these empirical approaches  describe the global systematics
of the level densities through the nuclear periodic
 table \cite{ref:RTK97},
important nuclear structure effects may be overlooked,
resulting in inaccurate level densities for 
at least some of the  nuclei.
Our present calculations seem to reproduce  available experimental
 densities with good accuracy (typically within a factor of 2)  for
$E_x\alt 20$~MeV.
Consequently, our extracted level density parameters are expected
 to be more accurate than the empirical ones.
Our values for  $a$ and $\Delta$ are obtained by fitting the BBF
 to the SMMC results in the energy range  $4<E_x<22$~MeV.

\begin{figure}
\epsfysize=7.0 cm
\centerline{\epsffile{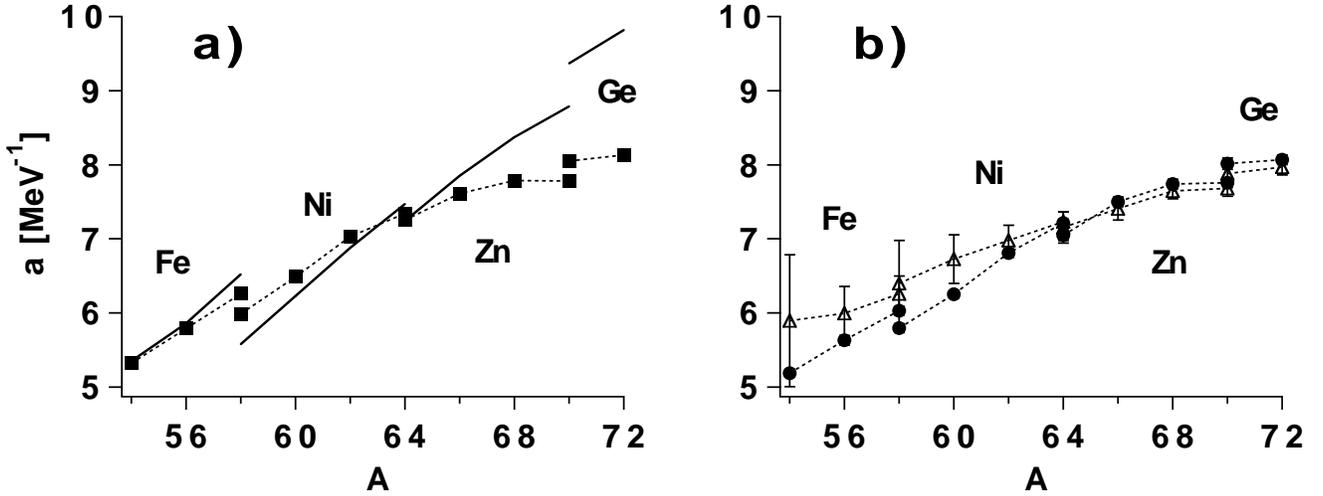}}
\vspace{2mm}
\caption{Single-particle level density parameter $a$
as a function of $A$, obtained from fitting the SMMC level densities
to the BBF (\protect\ref{BBF}).
Isotopes are connected by dotted lines.
Left:  $a$ for the total level densities (squares) in comparison
with the values obtained from the empirical formula of
Ref.~\protect\cite{ref:WFHZ} (solid lines).
Right: $a_+$  (circles) and $a_-$   (triangles) of the positive- and
negative-parity SMMC level densities, respectively.}
\label{fig:syst-a}
\end{figure}

\begin{figure}
\epsfysize=7.0 cm
\centerline{\epsffile{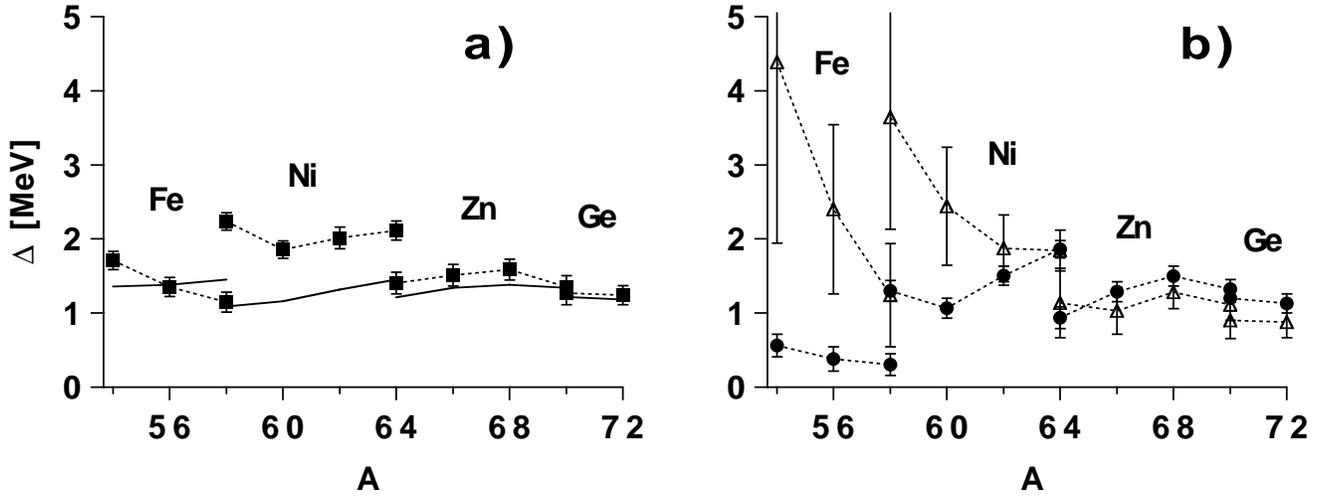}}
\vspace{2mm}
\caption{Backshift parameter $\Delta$
as a function of $A$, obtained from  SMMC calculations for total (left)
 and parity-projected (right) level densities.
Conventions are as in Fig.~\protect\ref{fig:syst-a}.}
\label{fig:syst-D}
\end{figure}

Figures~\ref{fig:syst-a} and \ref{fig:syst-D} show
the calculated values of $a$ and $\Delta$,
as well as the parity-dependent ones  $a_\pm$ and $\Delta_\pm$,
 as a function of $A$.
Isotopes are connected by dotted lines.  For comparison, 
the values obtained from the empirical formula of Ref.~\cite{ref:WFHZ}
 for the total level density parameters are  shown by solid lines.
Even-odd staggering in $\Delta$
 due to pairing correlations is not observed here 
since only even-even nuclei are considered.  It is interesting 
to determine whether  shell effects
can be observed in  the level density parameters.
Among the nuclei studied,
$^{54}$Fe and the  Ni isotopes have  $f_{7/2}$-shell closure
for protons ($Z=28$) or neutrons ($N=28$), respectively.
In the empirical values of the backshift parameter $\Delta$
of Ref.~\cite{ref:WFHZ}, no shell effects are seen for these nuclei.
In contrast, we find enhancement of $\Delta$ at $Z=28$ or $N=28$,
both for the total and parity-projected level densities.
Except for the $Z=28$ or $N=28$ nuclei,
the present $\Delta$ values for the total level densities
are close to those of Ref.~\cite{ref:WFHZ}.
On the other hand,  we do not observe shell effects
at $Z=28$ or $N=28$ in the
single-particle level density parameter $a$, which
 increases rather smoothly as a function of $A$.
Compared with the empirical estimates of Ref.~\cite{ref:WFHZ},
the present values of $a$  are not very different in the region $A\alt
65$,
but the $Z$-dependence within isobars tends to be weaker.
For $A>65$ our $a$ values are considerably smaller
than those of Ref.~\cite{ref:WFHZ}.
For  $^{68}$Zn, for example, we obtain $a=7.79$ as compared
 with the empirical value of $a=8.37$~\cite{ref:WFHZ}. Our value
 lies in between the experimental values of $a=7.25$ (assuming
half the rigid-body moment of inertia) and $a=7.97$ (assuming
the rigid-body moment of inertia)~\cite{ref:Dilg},
and is closer to the rigid-body value.
We shall return later  to the systematics of $a$.

We turn next to the parity-dependence of the level densities.
For $A\alt 60$ nuclei we observe that the level density parameters
are different for positive- and for negative-parity levels;
while $a_+$ is close to  $a$ of the total level
density, we find that $a_-$ is larger than $a_+$.
As $A$ increases, $a_-$ approaches $a_+$ (and therefore  $a$).
Similarly, for $A\alt 60$ nuclei $\Delta_-$ is substantially larger than
 $\Delta_+$ and both $\Delta_\pm$ are different from $\Delta$.
 The difference between $\Delta_+$ and $\Delta_-$ becomes smaller
as $A$ increases.
The observed  parity-dependence originates in the subshell structure
in this mass region,
where negative-parity states in even-even nuclei are possible
only when the $g_{9/2}$ level is populated.
Because of  the energy gap between the $pf$ and $g_{9/2}$ orbits
we expect the negative-parity level density to be lower
than the positive-parity level density at low energies.
Thus the backshift $\Delta_-$ should be larger than $\Delta_+$.
On the other hand, at high excitation energies
  positive- and negative-parity level densities are approximately equal,
as in the Fermi gas model.
Therefore, in the low energy region the negative-parity density is
 expected to rise more quickly
as a function of energy~\footnote{
Our BBF parametrization
 of the parity-dependent level densities is valid only
 for $E_x  \alt 20$ MeV.
At higher energies $\rho_+ \simeq \rho_-$ and therefore
 $a_+ \simeq a_-$ should hold.}
, i.e. $a_- > a_+$.
As $A$ increases, excitations (especially of neutrons)
to the $g_{9/2}$ orbit become easier,
 lessening the difference between $\rho_+$ and $\rho_-$.
For $A\agt 65$ we find that the values of  $a_\pm$ and $\Delta_\pm$
are close to those of the total level densities.

\begin{figure}
\epsfysize=7.0 cm
\centerline{\epsffile{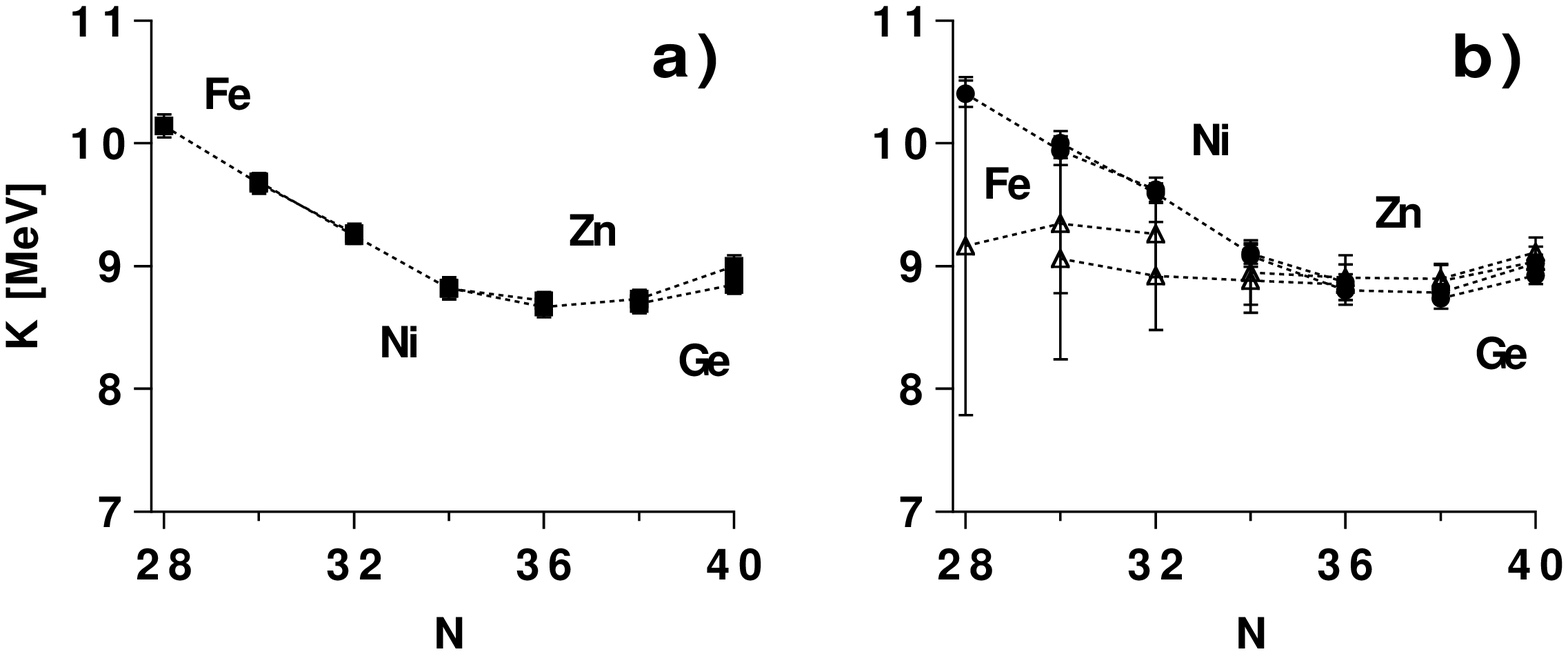}}
\vspace{2mm}
\caption{$K\equiv A/a$  as a function of  neutron number $N$
from the SMMC calculations for total (left) and parity-projected
 (right) level densities.
Isotopes are connected by dotted lines.}
\label{fig:syst-K}
\end{figure}

In the conventional Fermi gas model
the single-particle level density parameter $a$
is predicted to be proportional to $A$~\cite{ref:BM1}.
To investigate the $A$-dependence of $a$, it is customary to define
the quantity $K\equiv A/a$.
In the Fermi gas model $K$ is nearly constant ($\sim 16$~MeV).
In the empirical model of Ref.~\cite{ref:WFHZ},
 $K$ decreases nearly linearly as $A$ increases 
within a family of isotopes,
i.e. the level density increases more rapidly as a function
 of $E_x$ for heavier isotopes.
On the other hand, the empirical formula of Ref.~\cite{ref:Dilg}
(see Eq.~(10) of \cite{ref:Dilg}) predicts a gradual increase of $K$.
In the present microscopic SMMC calculations,
we find that $K$  (for the total level densities)
depends smoothly on the neutron number $N$, and is almost independent
of the proton number $Z$ as is shown in Fig.~\ref{fig:syst-K}.
$K \sim 10$~MeV at $N=28$, and decreases
  towards the middle of the $N=28-50$ shell to a value
 of $\sim 8.5$~MeV.
This behavior clarifies the shell systematics of $a$.
A similar behavior is observed  for $K_+\equiv A/a_+$.
Although  we cannot  make definite conclusions because of the
 large statistical errors,  $K_-\equiv A/a_-$ seems to be 
roughly constant.

In conclusion,  using the shell model Monte Carlo method
in the complete $(pf+0g_{9/2})$-shell, we have calculated 
microscopically total and parity-projected level densities  
for $\beta$-stable even-even nuclei
in the Fe to Ge region. We  have studied the systematics and shell
effects in both the single-particle level density parameter
$a$ (as well as
 the parameter $K=A/a$) and the backshift $\Delta$.
The shell structure observed in $\Delta$ at $Z=28$ or $N=28$
 and the systematics of $K$
differ from those given by various empirical formulae.

This work was supported in part by the  DOE grant
DE-FG-0291-ER-40608,
and by the Ministry of Education, Science and Culture of Japan
(grant  08740190).
Computational cycles were provided  by the IBM SP2 at JAERI and
Fujitsu VPP500 at RIKEN.

\clearpage

\begin{thebibliography}{99}
\bibitem{ref:Dilg} W. Dilg, W. Schantl, H. Vonach and M. Uhl,
Nucl. Phys. {\bf A217} (1973) 269.
\bibitem{ref:Hui}  J. R. Huizenga, H. K. Vonach, A. A. Katsanos,
A. J. Gorski and C. J. Stephan, Phys. Rev. {\bf 182} (1969) 1149.
\bibitem{ref:LVH} C. C. Lu, L. C. Vaz, J. R. Huizenga,
Nucl. Phys. {\bf A190} (1972) 229.
\bibitem{ref:BM1} A. Bohr and B. R. Mottelson, {\it Nuclear Structure}
vol.~1 (Benjamin, New York, 1969).
\bibitem{ref:HWFZ} J. A. Holmes, S. E. Woosley, W. A. Fowler and B. A.
Zimmerman, Atom. Data and Nucl. Data Tables {\bf 18} (1976) 305.
\bibitem{ref:CTT} J. J. Cowan, F.-K. Thielemann and J. W. Truran,
 Phys. Rep. {\bf 208} (1991) 267.
\bibitem{ref:WFHZ} S. E. Woosley, W. A. Fowler, J. A. Holmes and B. A.
Zimmerman,
Atom. Data and Nucl. Data Tables {\bf 22} (1978) 371.
\bibitem{ref:RTK97} T. Rausher,  F.-K. Thielemann and  K.-L. Kratz,
Phys. Rev. C {\bf 56} (1997) 1613.
\bibitem{ref:NA97} H. Nakada and Y. Alhassid,
 Phys. Rev. Lett. {\bf 79} (1997) 2939.
\bibitem{ref:G88} S. M. Grimes, Phys. Rev. C {\bf 38} (1988) 2362.
\bibitem{ref:MF75} K. K. Mon and J. B. French,
 Ann. Phys. (N.Y.)  {\bf 95} (1975) 90.
\bibitem{ref:SG88} R. Strohmaier and S. M. Grimes,
 Z. Phys. A {\bf 329} (1988) 431.
\bibitem{ref:KM96} V. K. B. Kota and D. Majumdar,
 Nucl. Phys. A {\bf 604} (1996) 129.
\bibitem{ref:SMMC} G. H. Lang, C. W. Johnson, S. E. Koonin and
 W. E. Ormand, Phys. Rev. {\bf C48} (1993) 1518.
\bibitem{ref:ABDK} Y. Alhassid, G. F. Bertsch, D. J. Dean and
 S. E. Koonin,  Phys. Rev. Lett. {\bf 77} (1996) 1444.
\bibitem{DZ96}  M. Dufour and  A. P. Zuker, Phys. Rev. C {\bf  54}
 (1996) 1641.
\bibitem{ref:ADK} Y. Alhassid, D. J. Dean, S. E. Koonin, G. Lang,
 and W. E. Ormand, Phys. Rev. Lett. {\bf 72} (1994) 613.
\bibitem{ref:LG92} E. Y. Loh Jr. and J. E. Gubernatis, in
 {\it Electronic Phase Transitions}, edited by W. Hanke and
 Yu. V. Kopaev (North Holland, Amsterdam, 1992), p.~177.
\end{thebibliography}
\end{document}